\documentclass{article}
\usepackage[english]{babel}
\usepackage{amsfonts,amssymb, amsmath}

\begin{document}

\title{Comment on: "St\"{a}ckel and Eisenhart lifts, Haantjes geometry and Gravitation"}
\author{A.V. Tsiganov\\
St.Petersburg State University, St.Petersburg, Russia\\
e--mail: andrey.tsiganov@gmail.com}
\date{}
\maketitle

\begin{abstract}
One of the oldest methods for constructing integrable Hamiltonian systems, proposed by Jacobi, recently is being presented  as a novel St\"{a}ckel lift construction related with Haantjes geometry. It may cause some confusion.
 \end{abstract}

In \cite{temp} authors introduce the notion of Stäckel lift as a novel geometric setting for the construction of large classes of integrable Hamiltonian systems.The authors claim that this method  extends the geometric framework underlying both the Riemannian and the Lorentzian-type classical Eisenhart lifts and that this method is intimately related with Haantjes geometry. 

The Jacobi separation of variables method consists of two steps which were  described by Jacobi in his lectures \cite{jac}, on page 198 or 199:
\vskip0.2truecm
\textit{
 "Die Hauptschwierigkeit bei der Integration gegebener Differentialgleichungen scheint in der Ein\-f\"{u}h\-rung der richtigen Variablen zu bestehen, zu deren Auffindung es keine allgemeine Regel giebt. Man muss daher das umgekehrte Verfahren einschlagen und nach erlangter Kenntniss einer merkw\"{u}rdigen Substitution die Probleme aufsuchen, bei welchen dieselbe mit Gl\"{u}ck zu brauchen ist."}
 \vskip0.2truecm
 Because these Jacoby ideas had a profound influence on the further developments of separability theory
  we also present the English translation from the Arnol’d book  \cite{arn}, p. 266:
   \vskip0.2truecm
   \textit{
   “The main difficulty in integrating a given differential equation lies in {introducing} convenient variables, which there is no rule for finding. Therefore, we {must} travel the reverse path and after finding some notable substitution, look for problems to which it can be successfully applied.”}
   \vskip0.2truecm
So, on the first step we {have to find} canonical transformation from original variables to convenient variables $q=(q_1,\ldots,q_n)$ and $p=(p_1,\ldots,p_n)$ with only nonzero Poisson brackets $\{q_i,p_i\}=0$ that
allows us  to integrate a given Hamilton-Jacobi equation. 

We now have a complete description and classification of all Hamilton–Jacobi equations that are separable in orthogonal curvilinear coordinates in $\mathbb R^n$ and on $\mathbb S^n$. Furthermore, we have such convenient variables of separation for algebraically integrable systems, including integrable tops, geodesic flow on $SO(n)$, the Kac–van Moerbeke lattice, Toda systems, the Gross–Neveu system and the Kolossof potential, among others \cite{van}. 

These results build the foundation for the second step, in which we substitute these convenient
variables $(p,q)$  into some new separated relations 
\[
\phi_i(q_i,p_i,\alpha_n,\ldots,\alpha_n)=0\,,\quad i=1,\ldots,n \qquad\mbox{with}\qquad \det\left(\frac{\partial \phi_i}{\partial \alpha_j}\right)\neq0\,,
\]
solve these $n$ equations with respect to $n$ parameters $\alpha_1,\ldots,\alpha_n$ and obtain new compatible Hamilton-Jacobi equations
\[\alpha_k=H_k(q,p)\,,\qquad k=1,\ldots,n,
\]
where $H_1,\ldots, H_n$ are functionally independent Hamiltonians in the involution $\{H_i,H_j\}=0$, according to the Jacobi theorem on the Hamilton-Jacobi equation.

This well-known Jacobi construction of integrable systems can easily be generalised by adding $m$ new variables
$(q_{n+1},\ldots,q_{n+m})$, $(p_{n+1},\ldots,p_{n+m})$ and parameters $\beta_1,\ldots,\beta_m$
and solving the new system of separated relations
\[
\tilde{\phi}_k(q_k,p_k,\alpha_n,\ldots,\alpha_n,\beta_{1},\ldots,\beta_{m})=0\,,\qquad k=1,\ldots,n+m
\]
with respect to parameters. Integrability of the obtained system is a sequence of the Jacobi and Liouville theorems.

In particular we can take a sum of two types  relations
\begin{align*}
&\tilde{\phi}_i(q_i,p_i,\alpha_n,\ldots,\alpha_n,\beta_{1},\ldots,\beta_{m})=0\,,\qquad i=1,\ldots,n\,,\\
&\tilde{\phi}_j(q_j,p_j,\beta_{1},\ldots,\beta_{m})=0\,,\quad j=n+1,\ldots,n+m\,.
\end{align*}
In fact, new "lifted" integrable systems from \cite{temp} can be directly obtained by solving such types of algebraic equations on a computer  without using  the St\"{a}ckel integration of the Hamilton–Jacobi differential equations.

Between 1891 and 1897, St\"{a}ckel published a series of papers on the calculation of the periods of quasi-periodic motion for a particular class of Jacobi integrable systems, see \cite{st91,st93,st93a,st93b,st95,st97,st97a} and references within. All these articles have now been made digital, so they can be found online. Thanks to artificial intelligence, they can be translated into any language.

To calculate the periods St\"{a}ckel  integrated a system of ODE's
\[\left(\frac{\partial}{\partial q_i}W_i(q_i)\right)^2=U(q_i)+\sum_{j=1}^nS_{ij}(q_i)\alpha_j\,,\qquad i=1,\ldots,n,\]
associated with separated relations of the form
\[
\phi_i(q_i,p_i,\alpha)=p_i^2-U(q_i)-\sum_{k=1}^nS_{ik}(q_i)\alpha_k=0\,,\qquad i=1,\ldots,n,
\]
and used Abelian sums to define the desired characteristics of motion at least formally.  

For other known integrable systems, for instance for algebraically integrable systems  \cite{van}, separated relations have a form
\[\phi_i(q_i,p_i,\alpha)=\varphi_i(q_i,p_i)-\sum_{j=1}^nS_{ij}(q_i)\alpha_j=0\,,\qquad i=1,\ldots,n.\]
It is easy to get some integrable systems on the extended phase space solving certain relations of the form
\begin{align*}
&\tilde{\varphi}_i(q_i,p_i,\beta_1,\ldots,\beta_{m})-\sum_{k=1}^nS_{ik}(q_i)\alpha_k=0\,,\qquad i=1,\ldots,n\,,\\
&\tilde{\varphi}_j(q_j,p_j,\beta_{1},\ldots,\beta_{m})-\sum_{k=1}^m\tilde{S}_{jk}(q_j)\beta_k=0\,,\quad j=n+1,\ldots,n+m\,.
\end{align*}
In fact it is a "novel St\"{a}ckel lift" from \cite{temp}. As mentioned above the Liouville integrability of the obtained Hamiltonian systems is  well-known sequence of the Jacobi and Liouville theorems.

In summary, in \cite{temp} the authors discuss some partial realisations of the well-known Jacobi idea. Since new integrable systems can be obtained by solving systems of algebraic equations of motion on a computer, many such systems can be produced every day. 

I don't find this way of building integrable systems interesting. I also don't think that Nijenhuis geometry and Haantjes algebra are so useful for this.

\end{document}